\documentclass[%
 reprint,
superscriptaddress,
 amsmath,amssymb,
 aps,
 prx
floatfix,
]{revtex4-2}

\usepackage[displaymath]{lineno}  
\usepackage{graphicx}
\usepackage{amssymb}
\usepackage{amsmath}
\usepackage[utf8]{inputenc}
\usepackage{cancel}
\usepackage[usenames,dvipsnames]{color}
\usepackage{psfrag}
\usepackage{epsfig}
\usepackage{bbm}
\usepackage{bm}
\usepackage{dsfont}
\usepackage{soul}
\usepackage{colonequals}
\usepackage{float}
\usepackage{enumitem}
\usepackage{appendix}
\usepackage{lmodern}
\usepackage[T1]{fontenc}
\usepackage{physics}                                                % Physics symbols
\usepackage{hyperref}
\usepackage[capitalize]{cleveref}
\usepackage{xcolor}			
\usepackage{siunitx} 
\usepackage{flafter}

\usepackage{graphicx}% Include figure files
\usepackage{hyperref}
\usepackage{dcolumn}% Align table columns on decimal point
\usepackage{bm}% bold math
\usepackage{xcolor}
\usepackage{siunitx}
\DeclareSIUnit\Molar{\textsc{m}}
\begin{document}

\preprint{APS/123-QED}

\title{Extracting work from hidden degrees of freedom}

\author{Lokesh Muruga}
\address{Faculty of Physics, University of Konstanz, Konstanz, Germany}
\author{Felix Ginot}
\address{Faculty of Physics, University of Konstanz, Konstanz, Germany}
\author{Sarah A. M. Loos}
 \email{sarah.loos@ds.mpg.de}
\address{Max Planck Institute for Dynamics and Self-Organization, Goettingen, Germany}
%\address{Max Planck Institute for Dynamics and Self-Organization, Goettingen, Germany}
\author{Clemens Bechinger}
%\address{Faculty of Physics, University of Konstanz, Konstanz, Germany}
 \email{clemens.bechinger@uni-konstanz.de}
\address{Faculty of Physics, University of Konstanz, Konstanz, Germany}
\address{Centre for the Advanced Study of Collective Behaviour, University of Konstanz, Germany
}


\newcommand{\sal}[1]{\textcolor{blue}{#1}}


\date{\today}% It is always \today, today,

\begin{abstract}
Thermodynamics establishes that information acquired through measurement can be converted into work, as exemplified by Maxwell’s demon and Szilard engines. Most experimental realizations of information engines, however, implicitly assume Markovian environments, in which information exchanged with the surroundings is irreversibly lost. Many physical systems instead exhibit environmental memory, with hidden degrees of freedom retaining correlations with the system’s past and giving rise to non Markovian dynamics. Whether and how such concealed memory can be harnessed as a thermodynamic resource has remained an open question. Here we experimentally demonstrate work extraction from environmental memory. Using time resolved measurements on an optically trapped Brownian particle in equilibrium, we implement a time delayed double measurement protocol that retrieves information via backflow from hidden bath degrees of freedom. We show that this information backflow alters relaxation dynamics, can be quantified independently of initial state effects, and when appropriately exploited enhances work extraction. Notably, we identify regimes in which the extracted work exceeds the energy stored in the observable degree of freedom alone. Our results establish environmental memory as an experimentally accessible thermodynamic resource and reveal how non Markovian dynamics can be systematically explored to improve the performance of information engines operating in time-correlated environments.
\end{abstract}

\maketitle

%%% ---------------------- InTroduction ---------------------- %%%

Maxwell’s famous \textit{Gedankenexperiment} illustrates that information about microscopic fluctuations 
of an equilibrium system can be converted into 
work extracted from a heat bath. The hypothetical demon thereby reveals a fundamental connection between information and energy conversion at small scales. 
This insight is formalized in information thermodynamics, which unifies work, entropy production, and information within a consistent nonequilibrium theory \cite{parrondo2015thermodynamics,cao2009thermodynamics,horowitz2010nonequilibrium,hasegawa2010generalization,esposito2011second,ito2013information,still2012thermodynamics}.

A cornerstone of information thermodynamics is a generalization of the second law of thermodynamics, first derived by Sagawa and Ueda, which bounds the average net work $W$ that can be extracted from a system during a measurement-based feedback protocol \cite{sagawa2010generalized,sagawa2012nonequilibrium}
\begin{align}\label{eq:Sagawa-Ueda-Rel}
W \le -\Delta F + k_\mathrm{B}T\,I(Z:M).
\end{align}
Here, $\Delta F$ is the free energy difference between the initial and final equilibrium states associated with the extraction protocol, and $T$ is the temperature of a weakly coupled heat bath~\cite{esposito2011second,seifert2016first}. The mutual information $I(Z:M)$ quantifies the average information gained about the microscopic system state $Z$ from a measurement with possible outcomes $M\in \Omega_M$. The state $Z$ represents the true physical state of the system, which fully determines its energetics and dynamics but is only partially accessible in practice. A measurement provides incomplete information about $Z$, reflecting finite resolution, noise, coarse graining, or unobservable degrees of freedom. 
Equation~\eqref{eq:Sagawa-Ueda-Rel} shows that information acquired through measurement constitutes a thermodynamic resource, enabling additional work extraction on average. 

Experimental realizations and theoretical studies of Szilard engines, tests of the Landauer bound~\cite{berut2012experimental,koski2014experimental,jun2014high,peterson2016experimental,ribezzi2019large,dago2021information}, and verifications of fluctuation theorems provided direct evidence of information-to-work conversion~\cite{toyabe2010experimental, ashida2014general, paneru2020colloidal, saha2021maximizing, still2022partially}. Most studies, however, assume Markovian, i.e. memoryless environments where exchanged information is irreversibly lost.
Often, however, this assumption breaks down: disordered materials, (bio)polymer solutions, periodically driven many-body systems, critical and active media but also open quantum mechanical systems typically exhibit pronounced temporal correlations that render the dynamics non-Markovian~\cite{SanchezDogic2012Nature, reichhardt2023reversible,furukawa2013nonequilibrium,yu2020hydrogels,admon2018experimental, liu2011experimental,horowitz2014thermodynamics}. Microscopically, such memory effects originate from hidden environmental degrees of freedom (\textit{h}DoF) that retain information about past system--environment interactions and thus influence subsequent dynamics and energetic exchanges.

The presence of \textit{h}DoF poses a central challenge for theoretical frameworks, as it complicates the interpretation of entropy production and the applicability of standard information–thermodynamic bounds~\cite{strasberg2019non,still2020thermodynamic,martins2021non,poulsen2022quantum,richter2024phase}. From the perspective of a Maxwell demon---or feedback controller---that relies solely on observed variables, the system appears partially opaque. Yet, memory constitutes \textit{information} about the system’s past retained in the environment, and such information has thermodynamic value. This raises the question of whether, and to what extent, environmental memory can be harnessed for work extraction.

In this work, we experimentally demonstrate work extraction from environmental memory using time-resolved position measurements of an optically trapped Brownian particle in a viscoelastic fluid. By extending conventional single-measurement schemes to time-delayed measurement pairs, we retrieve information stored in \textit{h}DoF and convert it into work. This approach enables a quantitative characterization of information backflow and establishes environmental memory as an experimentally accessible thermodynamic resource for enhancing the performance of information engines.

%%% ---------------------- Information backflow ---------------------- %%%

\begin{figure*}[ht]
\centering\includegraphics[width=0.95\linewidth]{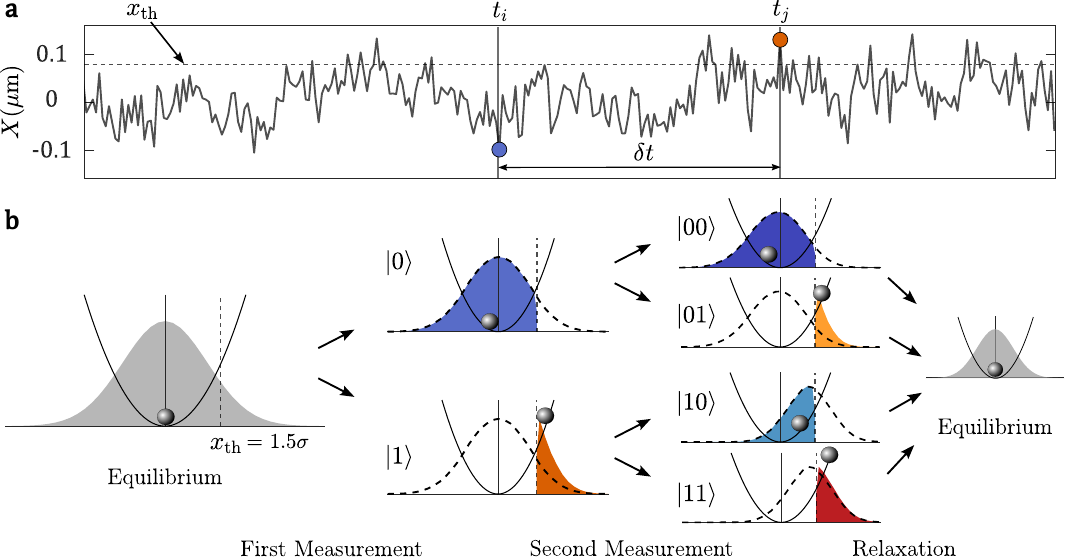}
\caption{\label{Prob_dist} {\textbf{Two-measurement protocol and state-conditioned particle distributions.}  
\textbf{a}, Measurement protocol illustrated using an experimental particle trajectory $x(t)$ (gray) corresponding to the state $|01\rangle$. At $t=t_i$ and $t_j=t_i+\delta t$, the particle position is compared to a fixed threshold $x_{\mathrm{th}}$; positions below and above the threshold are indicated by the blue and orange symbols and define the state $|0\rangle$ and $|1\rangle$, respectively. 
\textbf{b},  (left) Schematic equilibrium position distribution (gray) in the harmonic potential $V(x)$ (solid black); the vertical dashed line marks the threshold $x_{\mathrm{th}}$. (middle) The first measurement truncates the equilibrium distribution and prepares the conditional states $\rho_{|0\rangle}(x,t_i)$ (blue) and $\rho_{|1\rangle}(x,t_i)$ (orange), shown relative to the equilibrium distribution (black dashed). 
(right) A second measurement at $t_j=t_i+\delta t$ prepares a further refined conditional ensemble with the four initial states $|00\rangle$, $|01\rangle$, $|10\rangle$, and $|11\rangle$ and corresponding distributions $\rho_{|ij\rangle}(x,t_j)$ (dark blue, light orange, light blue, and red), which subsequently relax back to equilibrium (gray).
}
}
\end{figure*}

To elucidate the role of memory in work extraction, it is instructive to revisit Eq.~\eqref{eq:Sagawa-Ueda-Rel}. If the system's state $Z$ is taken to comprise the observable degree of freedom $X$ together with all environmental \textit{h}DoF $Y$, the dynamics becomes Markovian in the extended state space $Z=\{X,Y\}$, and Eq.~\eqref{eq:Sagawa-Ueda-Rel} formally applies with the mutual information $I((X,Y):M)$. Throughout, we assume isothermal conditions and restrict to protocols with $\Delta F=0$ (which is the case of any cyclic machine), allowing us to focus exclusively on information-driven work extraction.

It can be generally shown (see Methods) that $I((X,Y):M)=I(X:M)$, implying that a single measurement cannot access environmental memory. Consequently, Eq.~\eqref{eq:Sagawa-Ueda-Rel} remains unchanged as compared to the Markovian case, despite the presence of hidden \textit{h}DoF.

Further insight is obtained by considering work extraction conditioned on a specific measurement outcome $M=m$. The measurement transforms the equilibrium distribution $\rho_\mathrm{eq}(x)$ to the conditional distribution $\rho_m(x)=\rho(x|m)$, and the extractable work obeys \cite{esposito2011second,vaikuntanathan2011modeling,kawai2007dissipation}
\begin{align}\label{eq:second-law-KLD}
W_m \le k_\mathrm{B} T\, \mathcal{I}_m(Z) ,
\end{align}
with the relative entropy or Kullback--Leibler (KL) divergence $\mathcal{I}_m(Z)=D_\mathrm{KL}[\rho_m(z)||\rho_\mathrm{eq}(z)]$. Averaging this bound over $m$ recovers Eq.~\eqref{eq:Sagawa-Ueda-Rel}, as the mean relative entropy is the mutual information
%$\sum_mP_m I_m =I(x:m)$.
$\sum_{\Omega_M} P_m\mathcal{I}_m(Z)=I(Z:M)$. The outcome-resolved bound~\eqref{eq:second-law-KLD} immediately shows why \textit{h}DoF cannot be exploited through a single-time measurement: owing to conditional independence ($M \perp Y\,|\,X$), one has $D_\mathrm{KL}[\rho_m(x,y)||\rho_\mathrm{eq}(x,y)]=D_\mathrm{KL}[\rho_m(x)||\rho_\mathrm{eq}(x)]$.
This accords with physical intuition: memory manifests through temporal correlations, which are inaccessible to a single snapshot measurement. Exploiting memory therefore requires measurements at multiple times. 

Repeated measurements are generally beneficial in presence of measurement uncertainty (e.g., measurement error or binning), as they 
increase the available information about the system state. Consequently, even in Markovian systems, repeated measurements can enhance the amount of extractable work \eqref{eq:Sagawa-Ueda-Rel} \cite{ito2013information,horowitz2010nonequilibrium}. For protocols consisting of a measurement followed by feedback, however, the total extractable work from an initially equilibrium system is fundamentally bounded by the excess free energy in the post-measurement state relative to equilibrium \cite{parrondo2015thermodynamics}. 
In systems with memory, environmental \textit{h}DoF can contain additional excess free energy and lead to information backflow from the environment to the system, thereby, in principle, allowing work extraction beyond the Markovian limit (see Methods). 
%Owing to the associated information that flows back from the environment to the system, memory effectively raises the accessible information bounding the work , enabling, in principle, a larger amount of extractable work (see Methods).

\begin{figure*}[t]
\centering\includegraphics[width=0.95\linewidth]{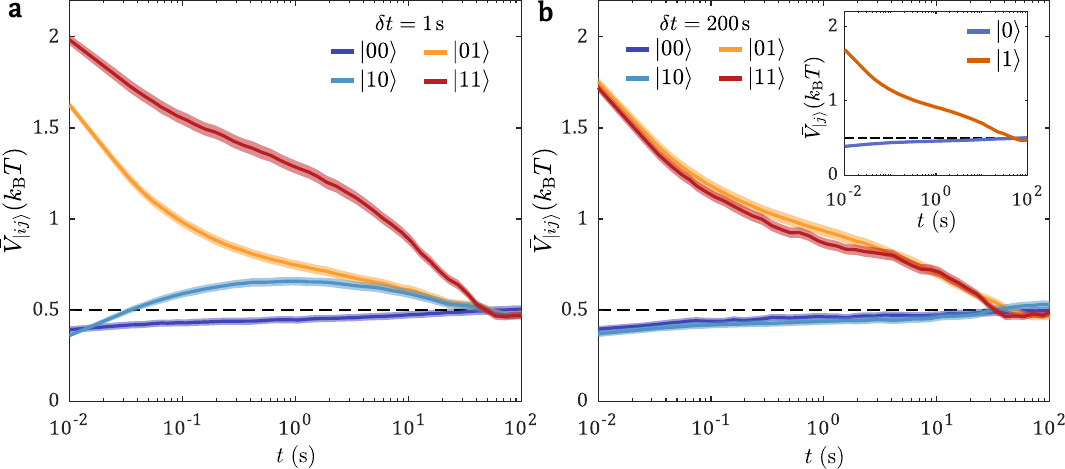}
\caption{\label{relaxation} {\textbf{Memory modifies relaxation: State-conditioned relaxation of the averaged particle potential energy.} 
\textbf{a}, Relaxation curves following a two-measurement protocol with $\delta t=\SI{1}{\second}$ for all possible conditional binary states which show marked differences.
\textbf{b}, With increasing $\delta t$, the influence of the first measurement progressively vanishes and the relaxation dynamics become determined solely by the second measurement, as demonstrated by the data obtained for $\delta t = \SI{200}{\second}$. This delay corresponds to the regime $\delta t \gg \tau_r$, for which the relaxation curves become indistinguishable from those following a single measurement (inset). Independent of $\delta t$, all curves asymptotically relax to the equilibrium energy $0.5\,k_{\mathrm{B}} T$ (dashed lines). The shaded bands correspond to the SEM calculated using a bootstrapping procedure (see Methods).
}
}
\end{figure*}

%%% ---------------------- Time-Resolved Measurements and Memory Effects ---------------------- %%%

As an experimentally accessible and well-controlled model system exhibiting environmental \textit{h}DoF, we study a colloidal silica particle of diameter \SI{2.7}{\micro\metre} suspended in a viscoelastic micellar solution composed of an equimolar \SI{8}{\milli\Molar} mixture of cetylpyridinium chloride monohydrate (CPyCl, 99\%) and sodium salicylate (NaSal, 99.5\%) (see Methods). At room temperature, the surfactants form a dynamical micellar network with a single structural relaxation time ~\cite{Spenley1993,Rehage1988}. For our concentrations, we yield a value of $\tau_r \approx \SI{6}{\second}$ (see Supplementary Section 2). As a result, a colloidal particle suspended in such solution exhibits pronounced non-Markovian behavior~\cite{baiesi_rise_2021,ginot2022barrier}. Because the micellar network dynamics is not directly observable in our experiments, it is considered as \textit{h}DoF storing information about past particle–bath interactions. The colloid is confined by optical tweezers in a harmonic potential $V(x)=\tfrac{1}{2}\kappa(x-\lambda)^2$, with $x$ the particle position, $\lambda$ the trap centre, and $\kappa$ the trap stiffness, respectively. Particle trajectories are recorded with rate $ \SI{100}{\per\second} $ and spatial resolution of $\pm\SI{5}{\nm}$ (see Fig.~\ref{Prob_dist}a). 

The particle position is sampled under equilibrium conditions, as seen by the stationary distribution shown in Fig.~\ref{Prob_dist}b. At each measurement time $t=t_i$, the position is compared to a threshold $x_{\mathrm{th}}$, assigning the state $m = |0\rangle$ for $X<x_{\mathrm{th}}$ and $m = |1\rangle$ otherwise. The temporal decay of each such single-measurement outcome is characterized by the corresponding conditional distribution $\rho_{|i\rangle}(x,t)$ (Fig.~\ref{Prob_dist}b). Note that a positional measurement does not alter the microscopic state of a classical system, rather it prepares a different statistical ensemble conditioned by the measurement outcome. Selecting a given outcome (e.g.\ $X > x_{\mathrm{th}}$) truncates the equilibrium Boltzmann distribution and creates a genuine nonequilibrium state (Fig.~\ref{Prob_dist}b). 

To access information about the full system, including \textit{h}DoF, we therefore perform a second measurement at $t_j = t_i + \delta t$. This defines a conditional state $|ij\rangle$ with possible outcomes $m \in \{ |00\rangle, |01\rangle, |10\rangle, |11\rangle\}$, characterized by conditional distributions $\rho_{|ij\rangle}(x,t)$ (Fig.~\ref{Prob_dist}b).
Such double-measurement selectively correlates the observed particle configuration with specific thermal excitations of \textit{h}DoF, which subsequently relax on their intrinsic timescale $\tau_r$. The coupled particle-network relaxation is indirectly reflected in the time evolution of the particle’s average potential energy,
\begin{equation}\label{def:pot-energy-measured-state}
   \bar{V}_{m}(t)=\int dx\, \rho_{m}(x,t)\,V(x).
\end{equation}

%\begin{equation}\label{def:pot-energy-measured-state}
 %  \langle V_{m} \rangle(t)=\int dx\, \rho_{m}(x,t)\,V(x).
%\end{equation}

The state-conditioned relaxation curves $\bar{V}_{|j\rangle}(t)$ and $\bar{V}_{|ij\rangle}(t)$, shown in Figs.~\ref{relaxation}, are obtained from $\SI{300}{\second}$ of trajectory data, averaged over $10^3$ realizations each. The relaxation from the single-measurement states $|0\rangle$ and $|1\rangle$ proceeds monotonically (inset Fig.~\ref{relaxation}b) toward the equilibrium  potential energy  $0.5\,k_{\mathrm{B}}T$, starting below and above this value, respectively.

Figure~\ref{relaxation} compares the energy relaxation for all possible measurement outcomes. The distinct relaxation curves demonstrate that a double measurement is sensitive to different states of the \textit{h}DoF. The difference between corresponding curves (e.g.\ $|01\rangle$ and $|11\rangle$) depends strongly on $\delta t$, confirming that the slow relaxation encodes memory of the system. As expected, these differences vanish in the limits $\delta t \to 0$ and $\delta t \to \infty$, consistent with the loss of correlations between successive measurements. Most notably, for the state $|10\rangle$ we observe a non-monotonic relaxation at intermediate delay times, where $\bar{V}_{|10\rangle}(t)$ initially increases before relaxing to equilibrium (Fig.~\ref{relaxation}a). This transient increase is an experimental signature of energy transfer from \textit{h}DoF to the particle. For large $\delta t$, the relaxation curves become purely monotonic, indicating that \textit{h}DoF equilibrated between measurements and no longer retains memory of the earlier state.

%%% ---------------------- Quantifying Information ---------------------- %%%
To experimentally quantify the additional information gained from a time-delayed second measurement, one might compare the conditional propagators $\rho_{|ij\rangle}(x,t)$ and $\rho_{|j\rangle}(x,t)$ for $t \ge 0$. Because these distributions already differ at $t=0$, such a comparison would conflate effects due to different initial conditions with genuine memory contributions. We therefore construct history-blind control distributions $\tilde{\rho}_{|ij\rangle}(x,t)$ by randomly subsampling trajectories from the single-measurement ensemble $\rho_{|j\rangle}(x,t)$ at $t=0$, such that the empirical distribution of particle positions at $t=0$ matches that of the corresponding two-measurement state (see Supplementary Section~3). By construction, this ensures
\begin{equation}
\tilde{\rho}_{|ij\rangle}(x,0) = \rho_{|ij\rangle}(x,0).
\end{equation}
For a memory-less system, $\tilde{\rho}_{|ij\rangle}(x,t) = \rho_{|ij\rangle}(x,t),$ for all $x$ and all $t\geq0$. 
Starting from such control ensembles, we then compare the averaged temporal evolution of observables obtained from single- and two-measurement outcomes, thereby isolating the contribution arising from information stored in \textit{h}DoF; which we refer to as information backflow. 
Any difference of the two ensembles also directly reflects information gained from the second, time-delayed measurement.

We characterize the time-dependent information content using the KL divergence with respect to the equilibrium distribution, $\mathcal{I}_{|ij\rangle}(t) = D_\mathrm{KL}[\rho_{|ij\rangle}(x,t)||\rho_{\mathrm{eq}}(x)]$.
 Applying the same construction to the control distribution yields the corresponding control information $\tilde{\mathcal{I}}_{|ij\rangle}(t)$. By construction, the two ensembles have identical initial information, $\tilde{\mathcal{I}}_{|ij\rangle}(0)=\mathcal{I}_{|ij\rangle}(0)$.

\begin{figure*}[ht]
\centering\includegraphics{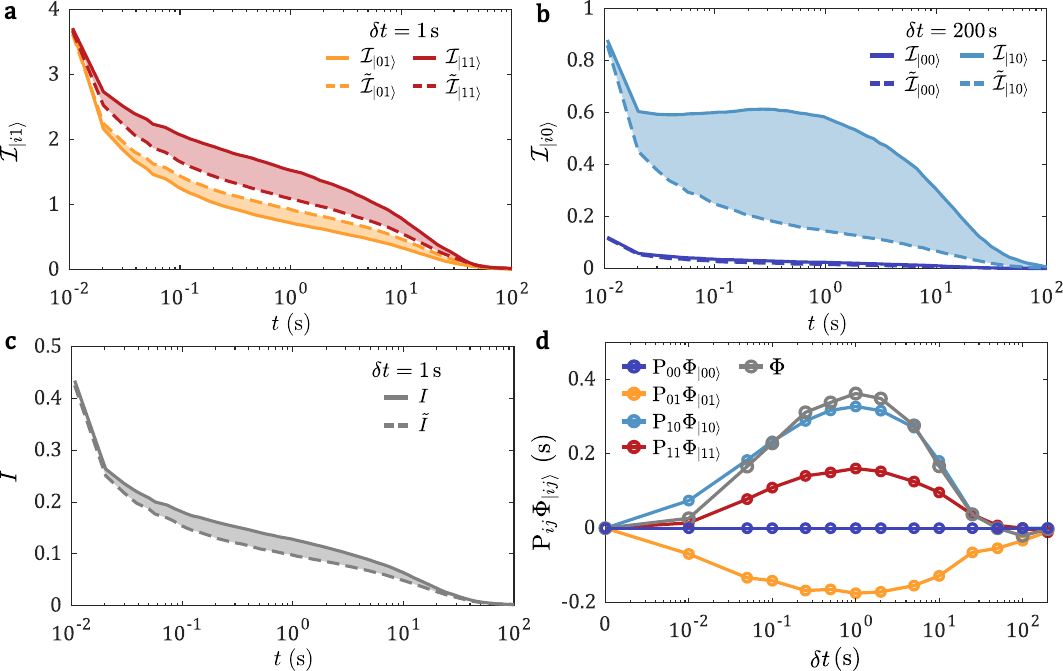}
\caption{\label{Information} \textbf{Information backflow: Information decay and non-Markovian order parameter.} 
\textbf{a}, Time evolution of the KL divergence $I_{|ij\rangle}(t)$ (solid lines) and $\tilde I_{|ij\rangle}(t)$ (dashed lines) for $\delta t = \SI{1}{\second}$, conditioned on the second measurement outcome $j = 1$. The  shaded regions denote enhanced persistence (accelerated decay) of information relative to the adjusted state. Notably, for the state $|01\rangle$, the difference $I_{|10\rangle}(t)-\tilde I_{|10\rangle}(t)$ is negative, corresponding to accelerated decay of information.
\textbf{b}, Same as a, but for states conditioned on $j = 0$. 
\textbf{c}, State-averaged information $I(t)$ and $\tilde I(t)$, obtained by weighting individual contributions with their empirical probabilities $\mathrm{P}_{ij}$. The shaded region highlights the net excess information retained at $\delta t = \SI{1}{\second}$.
\textbf{d}, Integrated order parameters weighted by their relative probabilities $\mathrm{P}_{ij} \Phi_{|ij\rangle}(\delta t)$ for individual measurement-conditioned states, together with the state-averaged quantity $\Phi(\delta t)$, showing a pronounced maximum at $\delta t \approx \SI{1}{\second}$.
}
\end{figure*}

In a Markovian system, the instantaneous state fully determines the future evolution, implying identical information decay for both ensembles (See Supplementary Section 4). The deviations observed at $\delta t \sim \SI{1}{\second}$ (Fig.~\ref{Information}a,b) therefore provide direct evidence for additional information stored in \textit{h}DoF of the micellar network, which is revealed only through the system’s evolution after the second, time-delayed measurement.

To obtain the average time-resolved information after a double-measurement, we weight each state $|ij\rangle$ by its probability $\mathrm{P}_{ij}$
\begin{equation}
I(t)=\sum_{ij} \mathrm{P}_{ij} \mathcal{I}_{|ij\rangle}(t)\,. 
\end{equation}
As shown in Fig.~\ref{Information}c, for $\delta t=\SI{1}{\second}$, $I(t)$ decays more slowly than the corresponding measure $\tilde{I}(t)$ despite identical initial values, demonstrating a history-dependent contribution beyond the instantaneous state.
Inspired by \cite{breuer2009measure}, we capture this excess contribution using the integrated order parameter
\begin{equation}
\Phi(\delta t)=\sum_{ij} \mathrm{P}_{ij}\int_0^{t}
 \big[\mathcal{I}_{|ij\rangle}(t)-\tilde{\mathcal{I}}_{|ij\rangle}(t)\big]\,dt,
\end{equation}
which quantifies the net information stored in \textit{h}DoF that influences the system’s subsequent dynamics. A value $\Phi(\delta t)\approx 0$ indicates memoryless (Markovian) behavior, for which the relaxation is fully determined by the instantaneous particle state. 
In contrast, $\Phi(\delta t)\neq 0$ indicates the influence of \textit{h}DoF, whose internal states affect the relaxation speed towards equilibrium. 
This interpretation also holds at the level of individual measurement-conditioned states and is reflected in the sign of $\mathcal{I}_{|ij\rangle}(t)-\tilde{\mathcal{I}}_{|ij\rangle}(t)$ (Figs.~\ref{Information}a-c). By varying the measurement interval $\delta t$, we thus map the temporal window over which memory controls relaxation speed of the particle dynamics. Importantly, a non-zero contribution to $\Phi(\delta t)$ corresponds to a delayed decay of information and coincides with regimes where energy is transiently transferred from \textit{h}DoF back to the particle, establishing a direct link between information backflow and energy backflow. Figure~\ref{Information}d shows the state-resolved contributions $\mathrm{P}_{ij}\Phi_{|ij\rangle}(\delta t) = \mathrm{P}_{ij}\int_0^{t}
\big[\mathcal{I}_{|ij\rangle}(t)-\tilde{\mathcal{I}}_{|ij\rangle}(t)\big]\,dt$. Despite pronounced differences in their shapes, the extreme values of all curves (except $\Phi_{00}$)  occur at almost the same delay time $\delta t \approx \SI{1}{\second}$. Consequently, also the state-averaged quantity
$\Phi(\delta t)$
exhibits a pronounced maximum at this delay, which corresponds to the timescale of information backflow of state $|10\rangle$ (Fig.~\ref{relaxation}a).

\begin{figure*}[ht]
\includegraphics{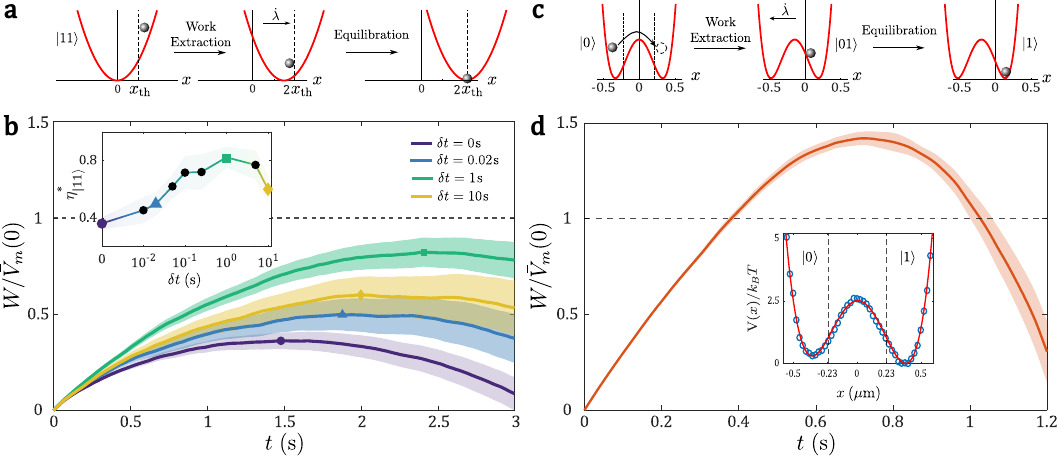}
\caption{\label{work_ext} \textbf{Work extraction from \textit{h}DoF through two-measurement protocols.}
\textbf{a}, Sketch of the work extraction protocol from a $|11\rangle$ state within a single well potential. After measuring a $|11\rangle$ state, the trap is translated at constant velocity $\dot{\lambda}$ during $t_\mathrm{f}=\SI{3}{\second}$ over a total distance $2x_\mathrm{th}$.
\textbf{b}, Averaged normalized work $W/\bar{V}_m(0)$ accumulated over time for increasing $\delta t$ (purple to yellow). Markers denote the maximal normalized work for each interval, ($\eta^*_{|11\rangle}$), which peaks at $\delta t = \SI{1}{\second}$ (see inset) but remains below unity, confirming that $W$ never exceeds $\bar{V}_m(0)$. Shaded bands represent the SEM from 50 events determined via bootstrap resampling (Methods).
\textbf{c}, Corresponding work extraction protocol for a double well potential with states $|0\rangle$ and $|1\rangle$ corresponding to the particle in the left and right well. After a barrier crossing event, i.e. a $|01\rangle$ state, the trap is translated at constant velocity during $t_\mathrm{f}=\SI{1.2}{\second}$ over a total distance of $\SI{360}{\nano\m}$.
\textbf{d}, Averaged normalized work $W/\bar{V}_m(0)$ extracted from a $|01\rangle$ state within a double well geometry. The maximum extracted work clearly exceeds $\bar{V}_m(0)$, demonstrating the extraction of energy stored within the \textit{h}DoF. The average has been obtained from 100 extraction protocols and the shaded area corresponds to the SEM obtained from bootstrapping (Methods).
}
\end{figure*}

%%% ---------------------- Work Extraction ---------------------- %%%
To test experimentally whether information retrieved from \textit{h}DoF can be converted into work, we compare to the fundamental information-theoretic bound. 
We use protocols of duration $t_\mathrm{f}$ consisting solely of rigid translational shifts of the external potential $V$. Specifying~\eqref{eq:second-law-KLD} to such conditions, we obtain for Markovian systems the looser bound,
\begin{align}\label{eq:energy-bound-trap-shifting}
W_m \leq \bar{V}_m(0) \,,
%\frac{\kappa}{2}\langle x\rangle_{m}^2 \leq \langle V\rangle_{m}\,.
\end{align}
where $\bar{V}_m(0)$ gives the mean energy of the post-measurement state [Eq.~\eqref{def:pot-energy-measured-state}]. 
This quantity sets the energetic starting point for our work extraction protocols. The reason why \eqref{eq:energy-bound-trap-shifting} assumes a purely energetic 
form is that rigid translations of the potential are isentropic operations, leaving both the equilibrium free energy and the Shannon entropy of the system invariant (see Methods). In contrast, in systems with memory, the extractable work is not bounded by $\bar{V}_m(0)$, since \textit{h}DoF can store additional excess free energy generated by thermal excitations, which becomes accessible during feedback.

In our experiment, we first consider a particle confined in a single-well harmonic potential, which serves as a minimal and well-controlled reference setting. Here, the state $|11\rangle$ is of particular interest because it exhibits the longest relaxation time among all measurement outcomes $|ij\rangle$ (Fig.~\ref{relaxation}a), indicating strong coupling to slow hidden modes and a pronounced influence of environmental memory. In the following, we therefore focus on work extraction protocols initiated from the $|11\rangle$ state.

We conduct the following protocol. After the second measurement, at time $t = 0$, the optical trap is translated at constant velocity $\dot{\lambda} = \SI{46}{\nano\m\per\s}$ over a duration $t_\mathrm{f}=\SI{3}{\second}$ yielding a distance of $2x_\mathrm{th}$ (Fig.~\ref{work_ext}a). The fluctuating work performed during the extraction process is given by
\begin{equation}
w = - \int_0^{t_\mathrm{f}} \frac{\partial V(x)}{\partial \lambda}\,\dot{\lambda}\,\mathrm{d}t \,,
\end{equation}
where positive values correspond to work extracted from the system. Figure~\ref{work_ext}b shows the average extracted work after measurement outcomes $m$ which we denote $W_m(t_\mathrm{f})$, normalized by the average potential energy $\bar{V}_m(0)$ available at the onset of extraction. The optimal conversion efficiency ($\eta^{*}_{|11\rangle}(\delta t)$) indicated by markers in Fig.~\ref{work_ext}(b) exhibits a pronounced dependence on the measurement interval $\delta t$ and the maximal value is obtained for $\delta t \approx \SI{1}{\second}$ (see Fig.~\ref{work_ext}b inset), coinciding with the maximum of the non-Markovian order parameter $\Phi$. This demonstrates that information retrieved from the second, time-delayed measurement enhances work extraction. Nevertheless, in the single-well geometry the extracted work remains below $\bar{V}_m(0)$. 

In the single-well case, the particle relaxes directly toward equilibrium once feedback begins, limiting the accessible work to the free energy contained in the conditioned particle distribution. To extract additional energy stored in hidden modes, one must transiently impede particle relaxation while the bath remains out of equilibrium.
To demonstrate work extraction exceeding $\bar{V}_m(0)$ that can hence be directly attributed to the presence of memory, we perform additional experiments in a double-well potential (see Methods for experimental implementation details). After a barrier-crossing event, the particle’s relaxation toward its original position is strongly hindered by the energy barrier, whereas the \textit{h}DoF relax on their intrinsic timescale, not affected by the barrier. 
This timescale separation creates a prolonged window during which free energy flows from the environment to the particle~\cite{ginot2022barrier}. A feedback protocol operating on the particle timescale can harvest this transient energy influx.

We define the two states of the double-well potential as $|0\rangle$ and $|1\rangle$, corresponding to the particle occupying the left and right well, respectively. Experimentally, this classification is implemented by comparing the particle position to two thresholds placed symmetrically around the potential maximum (Fig.~\ref{work_ext}c, dashed vertical lines). Both threshold positions correspond to a particle’s potential energy of $\approx 0.9\,k_{\mathrm{B}}T$. A $|01\rangle$ state indicates a recent particle jump from left to right across the potential barrier. 

To increase the number of detectable events within the limited experimental data set, we allowed $\delta t$ to take values within the interval $[0,\SI{200}{\s}]$ and aggregated the corresponding events to improve statistical significance.
To account for a small asymmetry in the potential wells, only unidirectional transitions are selected for the subsequent extraction protocol. Upon detection of such a $|01\rangle$, the trap is translated at $\dot{\lambda} = \SI{300}{\nano\m\per\s}$ for a duration of $t_\mathrm{f}=\SI{1.2}{\second}$, the latter being comparable to the structural relaxation time of the micellar solution. Figure~\ref{work_ext}d shows the extracted work, normalized by the potential energy available at the onset of extraction. In contrast to the single-well geometry, this quantity exceeds unity, demonstrating that thermal excitations of \textit{h}DoF of a non-Markovian bath can be converted into mechanical work. 

Our results highlight a general route to identify and exploit \textit{h}DoF as thermodynamic resource in systems with intrinsic memory. In this context, time-delayed or repeated measurements provide an operational handle to infer statistical information about the state of \textit{h}DoF from partial observations. This enables memory-aware control strategies, such as adaptive feedback protocols that tailor control actions to the inferred bath state and allow accelerated relaxation toward equilibrium in time-correlated environments where conventional Markovian shortcuts fail \cite{martinez2016engineered}. Beyond control, double measurements also suggest an operational criterion for memory erasure %in the sense of the Landauer bound, 
by signaling when all information-bearing hidden modes have relaxed and the dynamics becomes history independent. Beyond classical systems, the storage and backflow of information in unobserved environmental modes is a central theme in the theory of open quantum systems, where non-Markovianity has been widely discussed \cite{liu2011experimental,horowitz2014thermodynamics}. Our findings thus point toward a unifying perspective in which hidden degrees of freedom—classical or quantum—can be systematically characterized and exploited for information-based control, equilibration, and energy conversion in time-correlated environments.

%%% ---------------------- Bibliography ---------------------- %%%

% Before the Methods
\bibliography{Bib}

%%% ---------------------- Methods and others ---------------------- %%%
\section*{Methods}

%\noindent \textbf{No harvesting of memory after single positional measurements.}\\

%\noindent\textbf{No harvesting of memory after single positional measurements.}\\
\noindent\textbf{Single measurement cannot harvest memory.}
Here, we discuss an abstract setup in which a work extraction stage starts immediately after one or more measurements completed at time $t=0$, with no further measurements performed during the extraction stage. 

Let the extended state space comprise the observable degrees of freedom $X$ together with all \textit{h}DoF $Y$, which are not directly measurable. All degrees of freedom are assumed to be at the same temperature so that the combined system is initially in equilibrium. 

In the extended state space $Z=\{X,Y\}$ the dynamics becomes Markovian, and the generalized second law of information thermodynamics given in Eq.~\eqref{eq:Sagawa-Ueda-Rel} formally applies with the mutual information $I((X,Y):M)=\iint \rho(x,y,m) \ln \frac{\rho(x,y,m)}{\rho(x,y)\rho(m)}\mathrm{d}x\mathrm{d}m$. 
We specify to cases where the equilibrium free energy difference vanishes, $\Delta F=0$, which is appropriate for rigid potential shifting, and any cyclic protocol. We therefore omit the free energy term in Eq.~\eqref{eq:Sagawa-Ueda-Rel} throughout, which thus takes the form $W \le  k_\mathrm{B}T\,I((X,Y):M)$.

The measurement apparatus is assumed to not be directly influenced by $Y$. In an information-theoretical language, the measurement outcome $M$ is generated via a channel $\rho(m\mid x)$ that depends only on $X$. Consequentially \cite{cover1999elements},
\begin{align}\label{conditional-independence}
    \rho(m\mid x,y)= \rho(m\mid x)\,,~~~
    \rho(y\mid x,m)= \rho(y\mid x)\,,
\end{align}
implying the conditional independence property $Y \perp M \mid X$: $\rho(y,m\mid x)=\rho(y\mid x)\rho(m\mid x)$ \cite{cover1999elements}. 
%Therefore, the joint PDF factorizes $\rho(x,y,m)= \rho(y\mid x) \rho(m\mid x) \rho(x)\,.$
Note that this holds for continuous as well as discretized (threshold, binary) snapshot measurements. 
It follows directly that
\begin{align}\label{eq:MI-Identity1}
&I(Y : M \mid X)  
\nonumber \\
%\iint \rho(y,m|x)\ln\frac{\rho(y,m|x)}{\rho(y|x)\rho(m|x)}=
&=\int dx\, \rho(x)\iint dy\,dm\,\rho(y,m\mid x)\ln\frac{\rho(y,m\mid x)}{\rho(y\mid x,m)\rho(m \mid x)}
\nonumber \\
&=
\int dx \,\rho(x)\,\iint\,dy\,dm\, \rho(y,m\mid x)\ln 1
=
0 \,,
\end{align}
where we used that $\rho(y,m\mid x)=\rho(y\mid m,x)\rho(m\mid x)$\,. Using the decomposition of the mutual-information: $I((X,Y):M)=I(X:M)+I(Y:M\mid X)$ together with \eqref{eq:MI-Identity1}, we obtain
\begin{align}\label{eq:MI-Identity2}
    I((X,Y) : M)=I(X : M) \,.
\end{align}
This implies that Eq.~\eqref{eq:Sagawa-Ueda-Rel} reduces to the same form as for a Markovian system with no \textit{h}DoF.

%
%Note that the conditional independence likewise implies the Data Processing Inequality (DPI) \begin{align}\label{eq:MI-Identity3}I(X : M)\geq I(Y: M) \,.\end{align}

Now we turn to work extraction after a specific measurement outcome $m$. The outcome-resolved KL bound~\eqref{eq:second-law-KLD} reads $W_m \le k_\mathrm{B} T \,\mathcal{I}_m(X,Y),$
where $\mathcal{I}_m(X,Y)$ denotes the KL divergence $\mathcal{I}_m(X,Y)=D_\mathrm{KL}[\rho_m(x,y)||\rho_\mathrm{eq}(x,y)]$ giving the nonequilibrium free energy distance of the post-measurement state from equilibrium. We recall the notation, $\rho_m(x):=\rho(x\mid m)$.

Owing to the conditional independence $M \perp Y\,|\,X$, the posterior over the full system factors as
$\rho(x,y\mid m)=\rho(y\mid x)\,\rho(x\mid m)\,.$ Consequentially, for any realized $m$,
\begin{align*}
&D_{\mathrm{KL}}\left[\rho_m(x,y)\,\|\,\rho(x,y)\right]
\nonumber \\
&= \iint\,dy\,dx \rho(y\mid x)\,\rho(x\mid m)
    \ln\frac{\rho(y\mid x)\,\rho(x\mid m)}{\rho(y\mid x)\,\rho(x)} \\
&= \int\,dx \rho(x\mid m)\ln\frac{\rho(x\mid m)}{\rho(x)}
= D_{\mathrm{KL}}\!\left[\rho_m(x)\,\|\,\rho(x)\right]\,.
\end{align*}
Thus, $\mathcal{I}_m(X,Y)=\mathcal{I}_m(X)$ and the \textit{h}DoF disappear from the outcome-resolved bound \eqref{eq:second-law-KLD}, too.\\

\noindent \textbf{Multiple measurements allow harvesting memory.}
Owing to measurement uncertainties from sensor noise or binning, multiple measurements generally increase the acquired information compared to a single measurement, even for Markovian systems \cite{ito2013information,horowitz2010nonequilibrium}. Restricting for simplicity to the case of two measurements, the generalized second law \eqref{eq:Sagawa-Ueda-Rel} then formally involves the informational term $I((X_2, Y_2):(M_1,M_2))$.
Here, $(M_i,X_i,Y_i)$ denote the random variables at the time of the first measurement $t=-\delta t$ for $i=1$, and at the second measurement $t=0$ for $i=2$, after which the extraction stage is initiated.
While $M_i$ depends directly only on $X_i$, with $i=1,2$ respectively, the hidden variable $Y_2$ contains information about the earlier state $X_1$ through memory. Thus, $Y_2$ and $M_1$ are generally not independent conditional on 
$X_2$. Therefore, the
conditional independence property breaks down for the case of repeated measurements.

Decomposing the mutual information, 
$I(X_2, Y_2 : M_1,M_2) = I(X_2  : M_1,M_2)+I( Y_2 : M_1,M_2|X_2)$
, the generalized second law can now be written as
\begin{align}\label{eq:generalized-second-law-non-Markovian-double-measurement_v1}
      \frac{W}{k_\mathrm{B} T} \leq I(X_2 : (M_1,M_2)) +I( Y_2 : (M_1,M_2)|X_2)\,,
\end{align}
consistent with Refs.~\cite{ito2013information}. This decomposition makes explicit that memory can enhance the extractable work by the additional gain $I( Y_2 : (M_1,M_2)|X_2)\geq 0$, which quantifies the information about the \textit{h}DoF implicitly revealed by the repeated measurements.

Analogously, for fixed measurement outcomes $m1,m2$, the outcome-resolved bound \eqref{eq:second-law-KLD} on the extractable work is generally increased as compared to Markovian systems. This follows from the monotonicity of relative entropy under marginalization~\cite{cover1999elements}, which implies
\begin{align}
    D_\mathrm{KL}[\rho_{m1,m2}(x,y)||\rho_\mathrm{eq}(x,y)] 
    %\nonumber \\
    \geq D_\mathrm{KL}[\rho_{m1,m2}(x)||\rho_\mathrm{eq}(x)] .
\end{align}

We conclude that repeated measurements can in principle increase the extractable work beyond the Markovian bound by revealing additional information stored in the memory. This enhancement occurs provided that the measurements effectively probe correlations residing in the hidden degrees of freedom \cite{horowitz2014thermodynamics}.\\

%\noindent\textbf{Work extraction by rigid potential shifting operations.}
\noindent\textbf{Work extraction by rigid potential shifting.} By Eq.~\eqref{eq:second-law-KLD}, for a fixed measurement outcome $m$, the maximal extractable work in an isothermal process is bounded by the KL divergence. For Markovian systems, the latter can equivalently be expressed as the decrease in nonequilibrium free energy to the final equilibrium state \cite{esposito2011second}
\begin{align} W_m &\leq k_\mathrm{B}T\,D_{KL}\left[\rho_m(x)\|\rho_\mathrm{eq}(x)\right]
\nonumber \\
&= \mathcal{F}_\mathrm{neq}[\rho_m(x);V]-F_\mathrm{eq}[V], \end{align} where $\rho_{\rm eq}(x)\propto e^{-V(x)/(k_\mathrm{B}T)}$, and the nonequilibrium free energy is given by $\mathcal{F}_\mathrm{neq}[\rho;U] =\langle U\rangle_{\rho}-k_\mathrm{B}T\,S(\rho)$ for a state $\rho$ under external potential $U$ \cite{parrondo2015thermodynamics}. 
This expresses that no more work can be extracted than the free energy created by conditioning on the measurement outcome.
Equality can be achieved for reversible (quasistatic) protocols.

Now we discuss the specific example of work extraction 
restricted to rigid translations of the externally controllable potential $V\geq 0$,
\begin{equation}
V(x)\to V^\Lambda(x)=V(x-\Lambda),
\end{equation}
without any change of shape of $V$. In this case, the partition function
\begin{equation}
\mathcal{Z}(\Lambda)
=
\int dx \, e^{-\beta V(x-\Lambda)}
=
\int du \, e^{-\beta V(u)}
\end{equation}
is independent of $\Lambda$, so that the free energy $F_{\mathrm{eq}}=- k_B T \ln \mathcal{Z}$ is also invariant.
Therefore, all accessible equilibrium states have the same equilibrium free energy. Therefore, the only way to maximize extracted work is to minimize the nonequilibrium free energy before relaxation. Thus, maximizing extracted work is equivalent to minimizing the nonequilibrium free energy over the allowed shift parameter $\Lambda$ \cite{esposito2011second},
\begin{align}
W_m^{\max}
&=
\mathcal{F}_\mathrm{neq}[\rho_m;V]
-
\min_{\Lambda}\mathcal{F}_\mathrm{neq}[\rho_m;V^\Lambda]
\nonumber \\
&\leq k_\mathrm{B}T\,D_{KL}\left[\rho_m(x)\|\rho_\mathrm{eq}(x)\right].
\end{align}
Note that the last inequality follows from the fact that for any potential $U$ one has
$\mathcal{F}_\mathrm{neq}[\rho;U]\ge F_\mathrm{eq}[U],$
and translations leave the partition function invariant,
$F_{\rm eq}[V^\Lambda]=F_{\rm eq}[V]$.
Thus, the KL bound is generally not saturated for rigid-translation operations, unless the post-measurement distribution coincides with an equilibrium distribution of a translated potential.

Since the entropy $S[\rho_m]$ is invariant under rigid potential translations,
we find that, for Markovian systems where $V$ is the full potential energy of the system,
\begin{equation}\label{eq:Energy-bound-translation-Markovian}
W_m\leq W_m^{\max}
=
\bar{V}_m
-
\min_{\Lambda}[
\bar{V}^\Lambda_m ]~\leq \bar{V}_m\,,
\end{equation}
consistent with Ref.~\cite{abreu2011extracting}.
%
%Thus, for Markovian systems, the optimal strategy for a given $m$ is to choose $z$ that minimizes the nonequilibrium free energy in the translated potential \begin{align} W^\mathrm{max}_m=\max_z\Big[F_{\rm neq}[\rho_m(x);V]-F_{\rm neq}[\rho_m(x);V^z]\Big]\nonumber \\\leq k_\mathrm{B}T\,D_{KL}\!\left[\rho_m(x)\,\|\,\rho_\mathrm{eq}(x)\right],\end{align}
%where the nonequilibrium free energy is given by $F_\mathrm{neq}[q;U] =\langle U\rangle_{q}-k_\mathrm{B}T\,S(q)$ \cite{parrondo2015thermodynamics}. Note that the last inequality follows from the fact that for any potential $U$ one has $F_\mathrm{neq}[\rho;U]\ge F_\mathrm{eq}[U],$ and translations leave the partition function invariant, $F_{\rm eq}[V^z]=F_{\rm eq}[V]$. Since the entropic contributions to the free energies remain invariant under rigid potential translations, the bound reduces to the simpler form
%\begin{equation}\label{eq:Energy-bound-translation-Markovian}
%W_m
%\leq  W^\mathrm{max}_m =\max_z [\bar{V}_m-\bar{V^z}_m]\leq \bar{V}_m.\end{equation}
For the last inequality, we have made use of the choice $V\geq 0$.

In contrast, in non-Markovian systems, the internal coupling between the visible degree of freedom and the hidden degrees of freedom introduces additional sources of free energy, so that the simple bound \eqref{eq:Energy-bound-translation-Markovian} no longer applies. 

\begin{table}[h]
\centering
\caption{Composition of micellar solutions (prepared in $104.4~\mathrm{g}$ of deionized water).}
\label{table}
\begin{tabular}{lcc}
\hline
Concentration & CPyCl (mg) & NaSal (mg) \\ \hline
$5~\mathrm{mM}$ & $179.0$ & $80.1$ \\
$8~\mathrm{mM}$ & $286.4$ & $128.1$ \\ \hline
\end{tabular}
\end{table}

\noindent \textbf{Sample preparation and optical trapping.} Viscoelastic micellar solutions were prepared by dissolving cetylpyridinium chloride monohydrate (CPyCl) and sodium salicylate (NaSal) in equimolar proportions in deionized water (see Table~\ref{table} for mass specifications). The mixtures were stirred at $45~^{\circ}\mathrm{C}$ for $24~\mathrm{h}$ and equilibrated at $25~^{\circ}\mathrm{C}$ prior to use. Silica tracer particles ($2.7~\mu\mathrm{m}$ diameter) were dispersed at low volume fraction in $100~\mu\mathrm{m}$ thick glass capillaries and imaged at $100~\mathrm{fps}$ using a CMOS camera (Basler a2A2600-64umBAS). A $532~\mathrm{nm}$ laser (Coherent Verdi-V2) was focused through a $100\times$ oil-immersion objective (Zeiss, $\mathrm{NA}=1.3$) to create the optical potential. Using an acousto-optic modulator (AA Opto-Electronic) which controls the laser intensity, we realize a trap stiffness ($\kappa$) ranging from $0.5$ to $40~\mu\mathrm{N/m}$. To create the double-well potential, the trap center was oscillated at $700~\mathrm{Hz}$ using a galvanometer mirror—a frequency significantly higher than the particle's corner frequency, ensuring the particle experienced a time-averaged conservative potential. This potential was parameterized as $V_{0}(x) = a(x-x_1)^2(x-x_2)^2 + b(x-x_1)^2 + c$, where $x_1 \approx -350~\mathrm{nm}$ and $x_2 \approx 375~\mathrm{nm}$. Precise thermal stability was maintained at $25~^{\circ}\mathrm{C}$ using an objective heater (Okolab).

\noindent\textbf{Feedback protocols and statistical analysis.} Feedback control was implemented by evaluating particle positions in real-time to trigger automated trap displacements via a piezoelectric stage (Piezoconcept LT3.300). In the single-well geometry, measurement outcomes were defined relative to a spatial threshold of $x_{\mathrm{th}} = 70\,\mathrm{nm}$, whereas in the double-well landscape ($5~\mathrm{mM}$), states were assigned using thresholds at $\pm 230\,\mathrm{nm}$ (approximately $70\%$ of the distance to the potential minima). To maximize work extraction, control sequences were triggered upon detection of states characterized by slow relaxation dynamics: the $|11\rangle$ state for single-well experiments and the $|01\rangle$ state for the double-well. For these events, the trap was translated at a constant velocity—$140~\mathrm{nm}$ at $\dot{\lambda} \simeq 47~\mathrm{nm\,s^{-1}}$ for the single-well and $360~\mathrm{nm}$ at $300~\mathrm{nm\,s^{-1}}$ for the double-well. 

Statistical uncertainties were quantified using bootstrap resampling to estimate the error in the mean. For the state-conditioned relaxation curves (Fig.~\ref{relaxation}), which were subsampled from long continuous trajectories, we generated $500$ bootstrap realizations by randomly selecting $10^3$ segments. For the work extraction data in Fig.\ref{work_ext}c and d, where each curve corresponds to a specific measurement delay $\delta t$, bootstrapping was performed by generating $2,000$ resampled datasets. Each dataset was constructed by selecting the maximum number of independent trajectories with replacement from the original ensemble. In both cases, the shaded bands represent the standard deviation of the bootstrap distribution, providing a robust estimate of the standard error of the mean.

\section*{Acknowledgments}
The authors thank Eric Lutz and Semih H\"uner for fruitful discussions. 
The authors made limited use of Large Language Models (LLMs) to proofread the text and enhance the clarity and flow of the manuscript. 
CB acknowledges financial support by the ERC Adv. Grant BRONEB (101141477) and LM is supported by a scholarship of the Alexander von Humboldt foundation.

\section*{Authors Contributions}

LM and FG designed the experiments, which were carried out and analyzed by LM. The theory was developed by SL. CB directed the project. All authors participated in writing the manuscript.
\newline

\section*{Competing interests}
The authors declare no competing interests.

\section*{Additional information}
Supplementary Information is available online:

\section*{Data availability}

The processed data used to generate the figures in this study are available on Zenodo at the following link: https://doi.org/10.5281/zenodo.18715334. Due to the large size and complexity of the raw experimental data, they are available from the corresponding author upon reasonable request.

\section*{Code availability}

The computer code developed for the data analysis presented in this work is available on Zenodo at the following link: https://doi.org/10.5281/zenodo.18715334.

\end{document}